\documentclass[prl,
reprint,
showpacs,aps]{revtex4-1}
\pdfoutput=1

\usepackage{times}
\usepackage{graphicx}
\usepackage{natbib}
\usepackage{color}

\def\Ket#1{\left|#1\right\rangle}

\begin{document}

\title{Highly Efficient Source for Indistinguishable Photons of Controlled Shape}
\author{Peter Nisbet}
\author{Jerome Dilley}
\author{Axel Kuhn}
\affiliation {University of Oxford, Clarendon Laboratory, Parks Road, Oxford, OX1 3PU, UK}
\date{\today}
\begin{abstract}
We demonstrate a straightforward implementation of a push-button like single-photon source which is based on a strongly coupled atom-cavity system. The device operates intermittently for periods of up to 100 $\mu$s, with single-photon repetition rates of 1.0 MHz and an efficiency of greater than 60 \%. Atoms are loaded into the cavity using an atomic fountain, with the upper turning point near the cavity's mode centre. This ensures long interaction times without any disturbances induced by trapping potentials. 
The latter is the key to reaching deterministic efficiencies as high as obtained in probabalistic photon-heralding schemes. The price to pay is the random loading of atoms into the cavity and the resulting intermittency. However, for all practical purposes, this has a negligible impact.
\end{abstract}
\pacs{03.67.-a, 32.80.Qk, 42.50.Dv, 42.50.Pq, 42.50.Ex, 42.65.Dr}
\maketitle

Due to the large number of possible applications in quantum information processing, networking, and cryptography, deterministic single-photon sources are of prime importance \cite{Kimble:2008if}. The ideal system is one capable  of producing narrowband and indistinguishable photons on demand. For easy networking the souce should also be able to absorb single photons, mapping the photonic qubit onto the source where it can be stored for later use \cite{Boozer:2007el}. Due to its deterministic nature and the controllable coupling of a static qubit to a flying qubit we have chosen to use an atom-cavity system based on  a vacuum stimulated Raman process (V-STIRAP) to produce single photons. For an overview of the entire field please see \cite{Kuhn:2010p2121} and references therein. Other approaches include ions and quantum dots in cavities \cite{Keller:2004go, Moreau:2001wj}, electromagnetically-induced transparency (EIT) \cite{Chaneliere:2005fu}, and heralded down conversion sources \cite{Mosley:2008ir}. 

To perfect neutral-atom cavity sources many techniques have been developed such as intra-cavity dipole traps \cite{Puppe:2007fx} and feedback control of the motion of single atoms \cite{Kubanek:2009gs}.  Whilst enormous strides have been made, it has come at the cost of great experimental complexity.  In addition to this complexity, the electric and magnetic fields used to create long term traps for single atoms distort the atomic levels, reducing the photon emission probability and introducing additional de-phasing. With precision spectroscopy of the atom-cavity system and complete control of the atom's position using blue 'anti-trapping' dipole traps,  it should be possible to overcome these distortions, although at the cost of yet further experimental complexity. We have taken a different approach and use untrapped atoms to circumvent these difficulties, and have demonstrated high efficiency photon production with an  atom-cavity interaction time which is 'long enough' for all practical purposes.

\begin{figure}
\centering
\includegraphics[width=7.5cm]{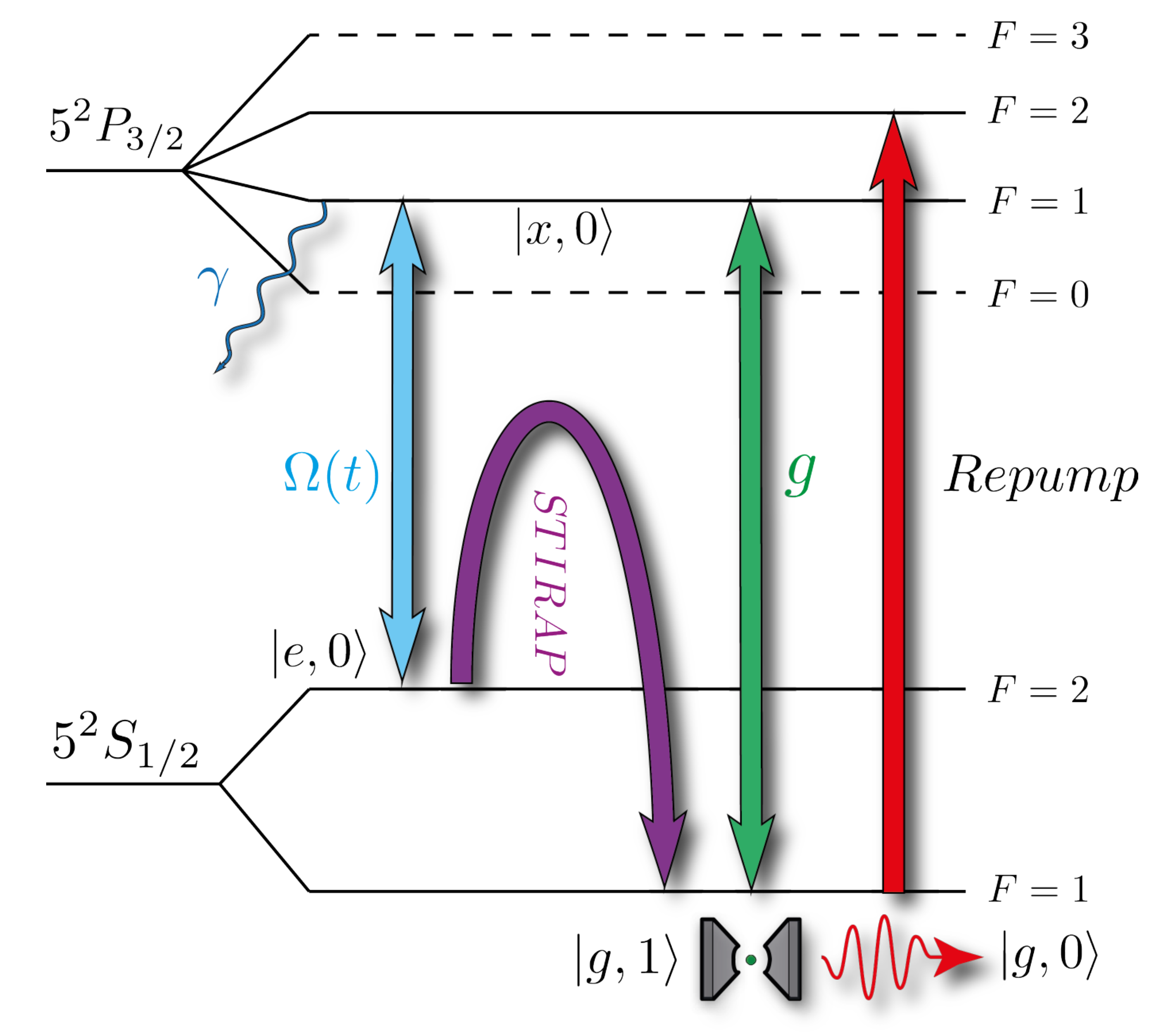}
\caption{Schematic diagram of the energy levels of the $^{87}$Rb$\,$D$_{2}$ line used for single photon production. The atomic states $\Ket{e},\Ket{x},\Ket{g}$ are involved in the STIRAP process and $\Ket{0},\Ket{1}$ denote the intra-cavity photon number.
\label{levels}}
\end{figure}

\begin{figure}[b]
\centering
\includegraphics[width=8.5cm]{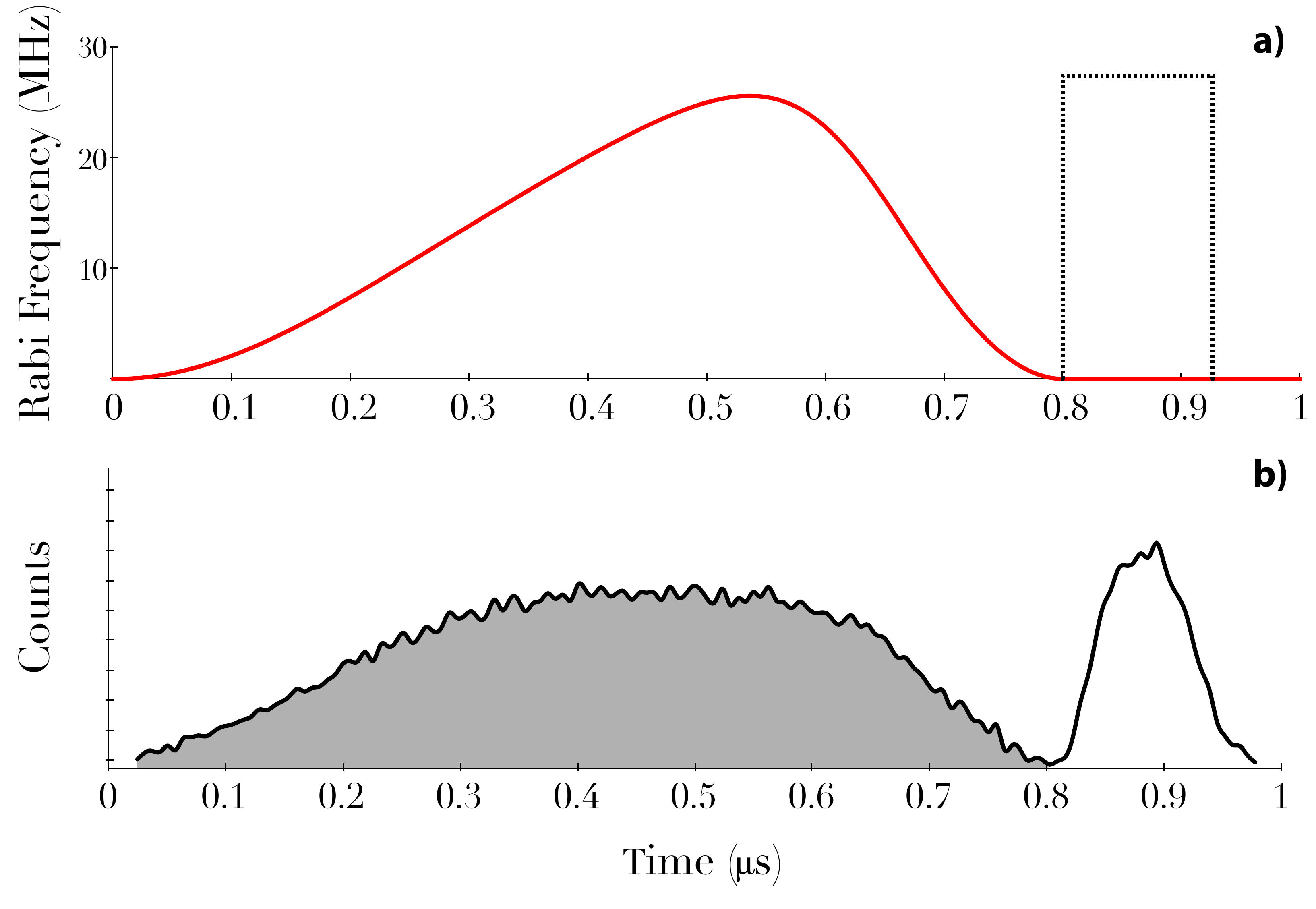}
\caption{(a) The sequence of driving pulse (solid) and re-pump pulse (dashed). The Rabi frequency of the re-pump pulse is not to scale. (b) Histogram of detector clicks, STIRAP photons are shared grey. The counts during the re-pump phase (unshaded) are mostly due to the beam clipping the cavity mirrors.}
\label{fig:pulses}
\end{figure}

Photons are produced using a V-STIRAP process \cite{Kuhn2002,Hennrich:2000p127}. Fig.\,\ref{levels} illustrates the level scheme for the $^{87}$Rb D$_{2}$ line ($5^{2}S_{1/2}\rightarrow 5^{2}P_{3/2}$).  A single atom located in a high finesse optical cavity with photon number $\Ket{n}=\Ket{0}$, is prepared in the $F=2$ hyperfine ground state denoted by $\Ket{e}$.  The cavity is resonant with the transition $F=1\rightarrow F'=1$,  $\Ket{g}\rightarrow\Ket{x}$, and the cavity vacuum state causes an electric dipole interaction with strength $g_{0}$.  The atom is driven with a laser on the $F=2\rightarrow F'=1$ transition with time dependent Rabi frequency $\Omega(t)$.  This combination of fields has the result of pumping the atom from $\Ket{e}\rightarrow\Ket{g}$ whilst creating a photon in the cavity mode,  $\Ket{e,0}\rightarrow\Ket{g,1}$.  The photon decays out of the cavity with rate 2$\kappa$.  Throughout the process the atom remains in a superposition of the two ground states $\Ket{e}$ and $\Ket{g}$ whilst the excited state $\Ket{x}$, which is subject to spontaneous emission $\gamma$, remains dark.
Once the system has decayed to the state $\Ket{g,0}$ it is decoupled from further evolution, and must be optically pumped to the state $\Ket{e,0}$ before another photon can be emitted.  This is achieved by a second laser pulse driving the atom from state $F=1\rightarrow F'=2$, from which it decays probabilistically to $F=2$.  The experimental sequence of pulses is shown in Fig.\,\ref{fig:pulses}.
    
One of the main challenges in CQED experiments is reliably coupling a single atom to an on-resonance cavity.  To reduce experimental complexity we have completely removed the intra-cavity traps commonly used to achieve this.  Instead we use a magneto-optical trap (MOT) and an atomic fountain to ballistically launch atoms into the cavity mode. The atom number is kept sufficiently low that the probability of two launched atoms entering the cavity mode at the same time is negligible, (only $0.26\%$ of all atom cavity interactions will occur with two atoms).  

\begin{figure}[!t]
\centering
\includegraphics[width=8.5cm]{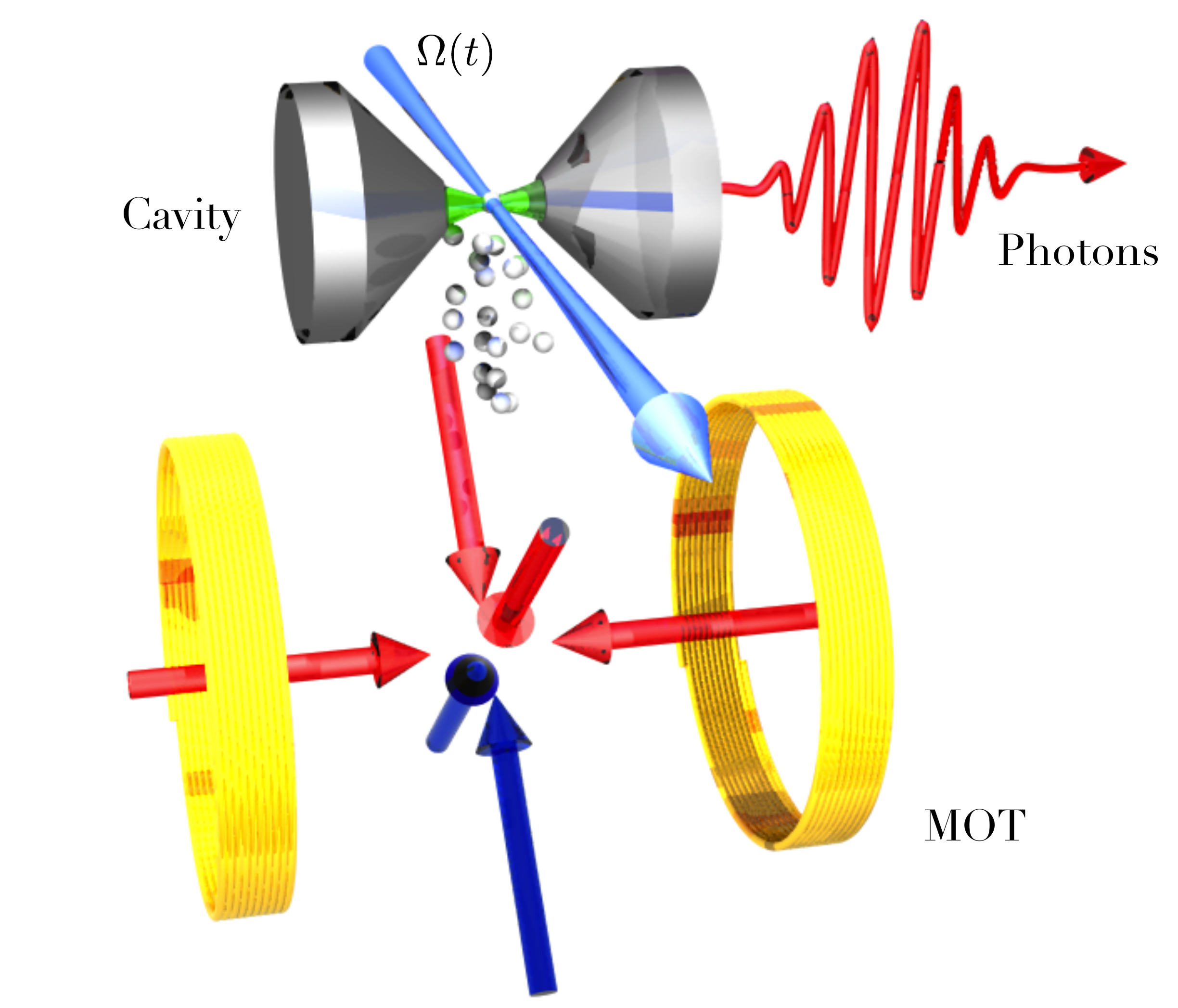}
\caption{Artist's view of a the arrangement of the cavity and MOT. The lower MOT beams are blue detuned relative to the upper to launch the atoms the $8\,$mm between the MOT and the cavity mode. Once in the cavity mode the atoms are driven by a Raman laser to produce single photons, colors correspond to those in Fig.\ref{levels}}
\label{fountain}
\end{figure}

Around $10^6$ atoms are prepared in a standard 6-beam MOT approximately $8\,$mm below the centre of the cavity mode Fig.\ref{fountain}.  This loading phase lasts for $\approx 75\,$ms and is assisted by UV light-induced desorption (LIAD), allowing for fast loading rates with a relatively low background pressure of $10^{-10}$ mbar. Following this, the MOT coils are switched off and the frequencies of the upper and lower molasses beams are detuned relative to each other. The atoms are cooled into a moving rest frame, the velocity of which is given by
\begin{equation}
v=\sqrt{2}\lambda\Delta f
\end{equation}
where $\Delta f$ describes the relative beam detuning between the upper and lower beams and $\lambda$ the laser wavelength. Fine control over the frequencies of these beams (tens of kHz) leads to intrinsically fine control over the launching velocity. Varying the velocity of the launch allows the throw to be tuned so that the turning point of the atomic motion is in the cavity mode. Using simple ballistic flight arguments, and assuming a cavity mode diameter of $d=40\,\mu$m one could, in theory, achieve maximum interaction times of  $t_{int}=2.\sqrt{2d/g}\approx4\,$ms. This however assumes that the atom perfectly traverses an anti-node of the cavity field with zero horizontal velocity.  The finite size and temperature of the atom cloud limits the achievable interaction time to hundreds of microseconds.

The cavity is constructed from two highly reflecting mirrors separated by a distance of $L=74\mu$m. A cavity finesse of $\mathcal{F}=85,000$ is achieved resulting in parameters of $(g,\kappa,\gamma)=2\pi \times (12,12,3)$ MHz, putting the atom-cavity system into the regime of strong coupling. 
The cavity was initially built with mirrors with $(T_1, T_2, L)= (40, <\!1,2)$ ppm, which resulted in a finesse of $\mathcal{F}>100,000$ and a large asymmetry so  that $96\%$ of photons emitted into the cavity could be collected from the same spatial mode.  The reduction in finesse occurred during the bakeout of the vacuum system.

The design of the cavity is intended to be both simple and inherently stable, whilst allowing for very good optical access. Both mirrors sit in ceramic mounts glued to shear piezo actuators (Noliac - CSAP03) which are glued to a non-magnetic stainless steel mount inside a UHV vacuum chamber. 
High passive stability is observed with several seconds required for the cavity frequency to drift by its HWHM $(12\,$MHz$)$. Active feedback is achieved using the Pound-Drever-Hall technique and PID regulator, allowing the cavity to be locked to the atomic resonance for many hours. The lock can be interrupted using a sample and hold circuit ($LF398$) during experimental runs: this allows the locking beam to be switched off while the passive stability ensures that the cavity remains on resonance for the $20\,$ms required for the atom cloud to pass through the cavity.

\begin{figure}
\centering
\includegraphics[width=8.5cm]{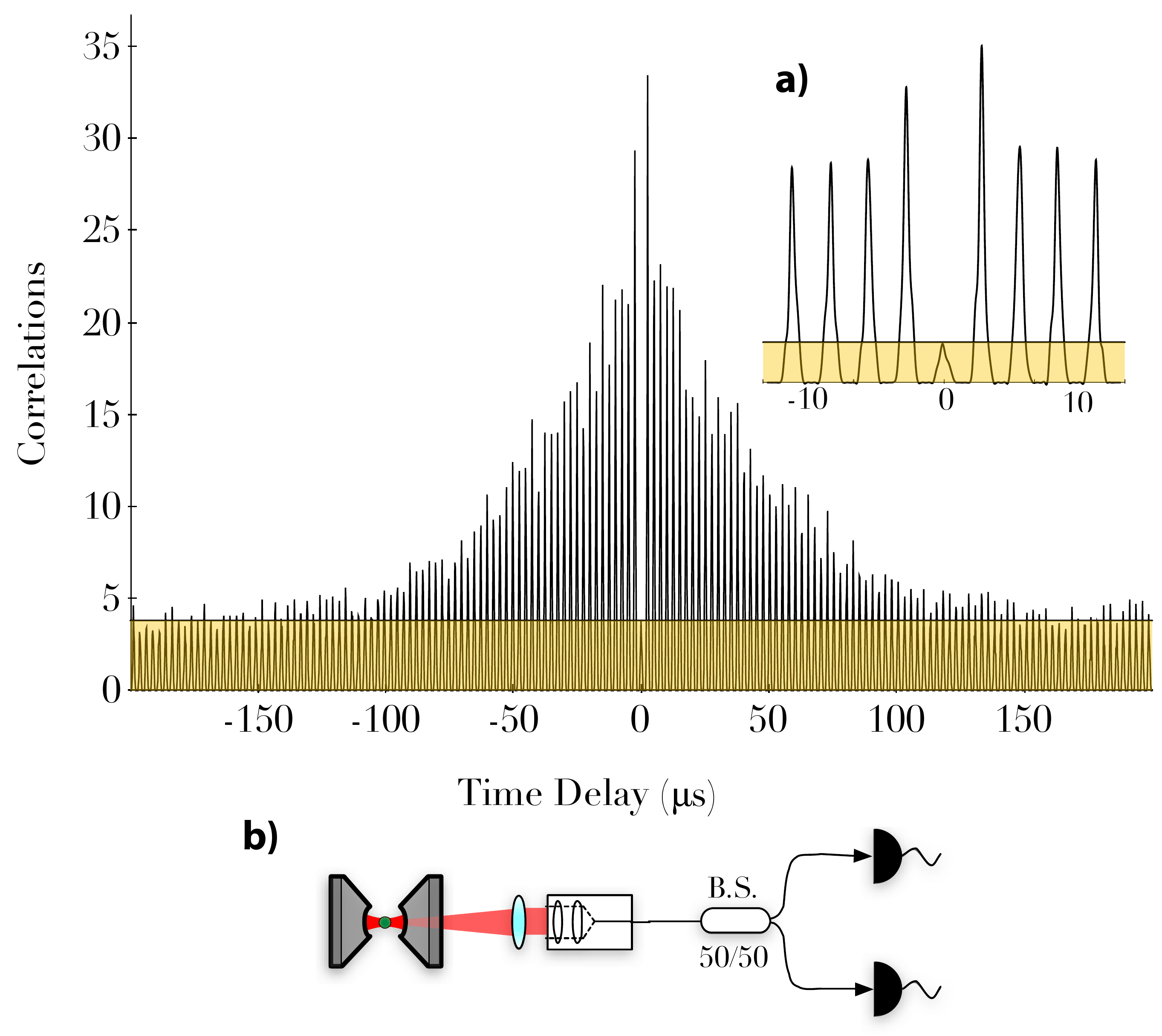}
\caption{Second order correlation function of the detected photons, the dark-count background level is shaded yellow. The missing central peak implies a single photon source, and the envelope of the peaks implies that the atom remains in the cavity for $100\,\mu$s. Inserts show a zoomed in view around $t=0$ (a) and the interferometer setup (b).}
\label{fig:g2}
\end{figure}

To verify that we are producing single photons the photon stream is interrogated using a Hanbury-Brown-Twiss type interferometer. A $50/50$ beam splitter is placed in the beam path and correlations between the clicks on each of the two output ports are measured.  This can be used to calculate the second order intensity correlation function of the emitted photons. The cross correlation of the photo-detectors D1 and D2 is shown in Fig. \ref{fig:g2}, which is defined as
\begin{equation}
g^{(2)}(\tau) = \frac{\langle P_{D1}(t) P_{D2}(t-\tau)\rangle}{(\langle P_{D1}(t)\rangle\langle P_{D2}(t)\rangle)},
\end{equation}
where $P_{D1}(t)$ and $P_{D2}(t)$ are the probabilities of detecting a photon at the corresponding detector \cite{Hennrich:740149}. This function includes both photon correlations and contributions from the dark noise of the detectors ($1\,$kHz) . Detector counts which occur during the optical re-pumping have been omitted, see Fig. \ref{fig:pulses} (b). This masking gives rise to the periodicity of the background which would otherwise be at a constant level (shown in yellow shading, the two atom contribution to this is negligible).
The $g^{(2)}(\tau)$ function exhibits the periodicity that one would expect for a pulsed source, with peaks separated by the repetition period. It can clearly be seen that the expected peak at $\tau=0$ is missing.  This demonstrates that there is only one photon produced per atom per pulse.  The envelope of the $g^{(2)}(\tau)$ correlation function is a consequence of the limited atom-cavity interaction time. At $100\,\mu$s this allows for the implementation of many interesting QIP schemes, e.g. entanglement generation, teleportation and gate operations \cite{Himsworth:2011hn,Wilk2007,Olmschenk:2009p1310}. 

\begin{figure}[!b]
\centering
\includegraphics[width=8.5cm]{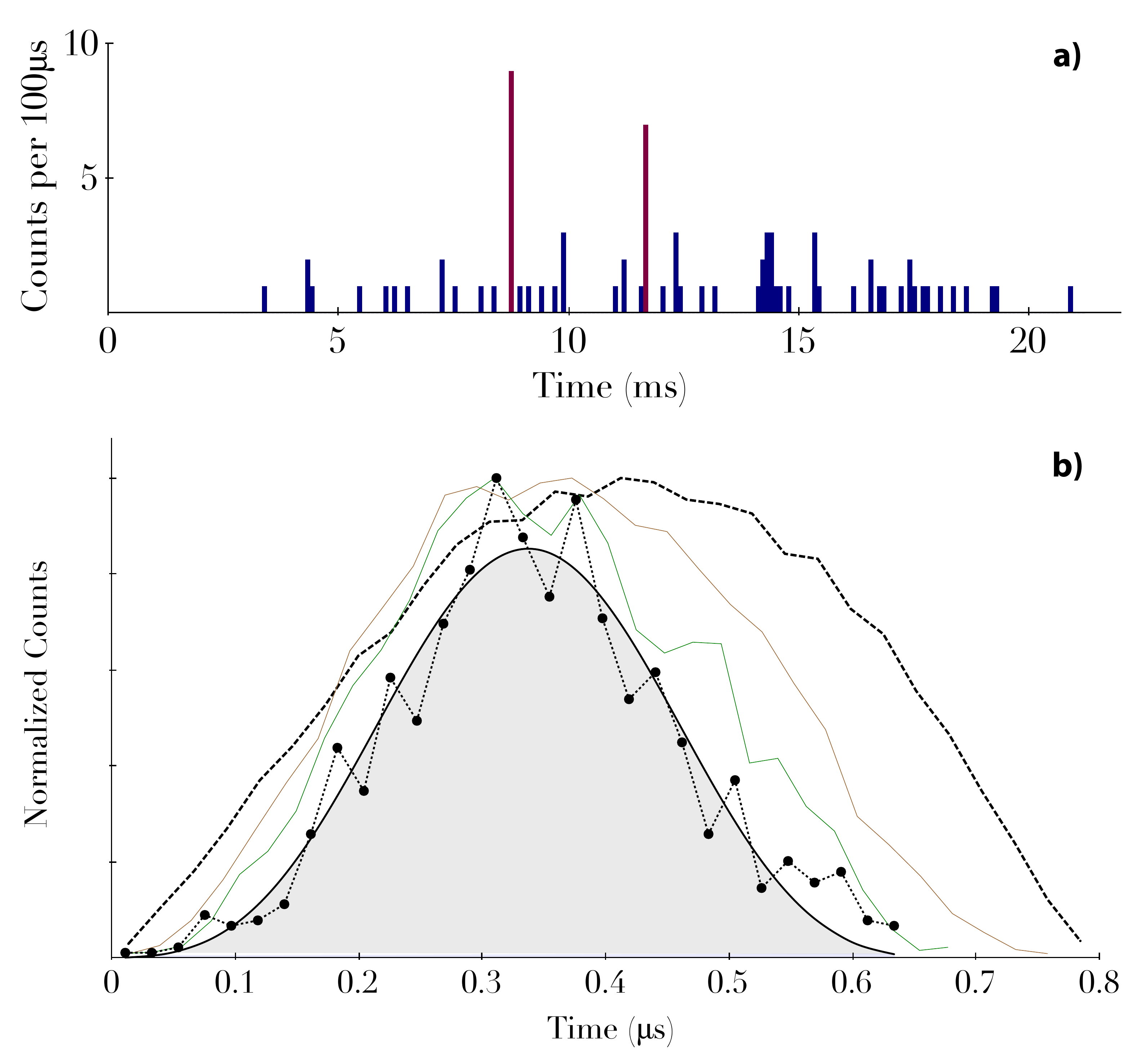}
\caption{(a) Histogram of photon arrival times, binned with the interaction time. Time bins in which an atom passes through the cavity are highlighted in red. By post-selecting these time bins the $|\psi(t)|^2=sin^4(t)$ shape (shaded) for a well coupled atom can be recovered (b).  The photon shapes are shown when not post-selecting (dashed), and selecting only photons which occur with more than 3 (brown), 5 ( green) and 7 (circle) counts per bin.    }
\label{fig:shape}
\end{figure}

A typical histogram of the photon arrival times is shown in Fig. \ref{fig:shape} (a). When an atom passes through the cavity the photon count rate increases sharply - this can clearly be seen by the red bars.  Post selection of these atom transit events allows for the efficiency of the photon production process for a well coupled atom to be determined.  The emission probability is calculated by conditioning on a detector click and then looking for clicks in the subsequent pulses.  The emission probability $P$ is observed to change as the atom moves through the cavity as the coupling $g$ depends on the atom's position within the mode.

As the most likely place for the atom to emit a photon is at the cavity center, calculating the efficiency by looking at the probabilty of emitting two sucessive photons will always underestimate $P$; the atom will have moved away from $g_{0}$ when the second emission takes place.
 Instead, we map the change in $P$ with multiple sucessive pulses which follows a gaussian curve.  It is possible to extrapolate this curve back to the origin to estimate the maximum single-photon emission probability $P_{max}$. 
 We find a detector click probability of  $15\pm2\%$. By including the detector quantum efficency of $70\,\%$ and a $65\,\%$ collection efficiency (coupling photons into to the fiber, optical losses and losses from the vacuum chamber viewport) this corresponds to a maximum photon emission probility of  $33\pm 3.5\,\%$.  As previously intimated, during the vacuum bakeout a dirty Rb dispenser caused the mirror losses to increase from $~2\,$ppm $\rightarrow18\,$ppm. This reduced the photon outcoupling efficiency from 96\% to only 50\%. Including this factor in the calculation gives the photon production probability inside the cavity, $P_{max}=66\pm 7\,\%$.

The pulse applied to the atom was calculated to produce a $\psi(t)=sin^2(t)$ shape using the method in \cite{Vasilev:2010ca}, having assumed a stationary atom experiencing the maximum coupling strength $g_{0}$. In reality, due to the imperfect positioning of the atom within the cavity mode, the atom will experience coupling strengths over the full range of $g=0\rightarrow g_{0}$. The grey shaded area in Fig. \ref{fig:pulses} (b) shows a histogram of the photon arrival times  from the beginning of the driving pulse (from a total of $10^5$ photon counts). This clearly deviates from an ideal $\psi(t)=sin^{2}(t)$ envelope and is due to photon emission from poorly coupled atoms.

The post selection used to determine the emission probability can also be used to recover the shape of the photons being produced. By selecting out only the photons from well coupled atoms the measured shape collapses to the expected $\psi(t)=sin^2(t)$, as shown in Fig. \ref{fig:shape} (b).

\begin{figure}[!b]
\centering
\includegraphics[width=8.5cm]{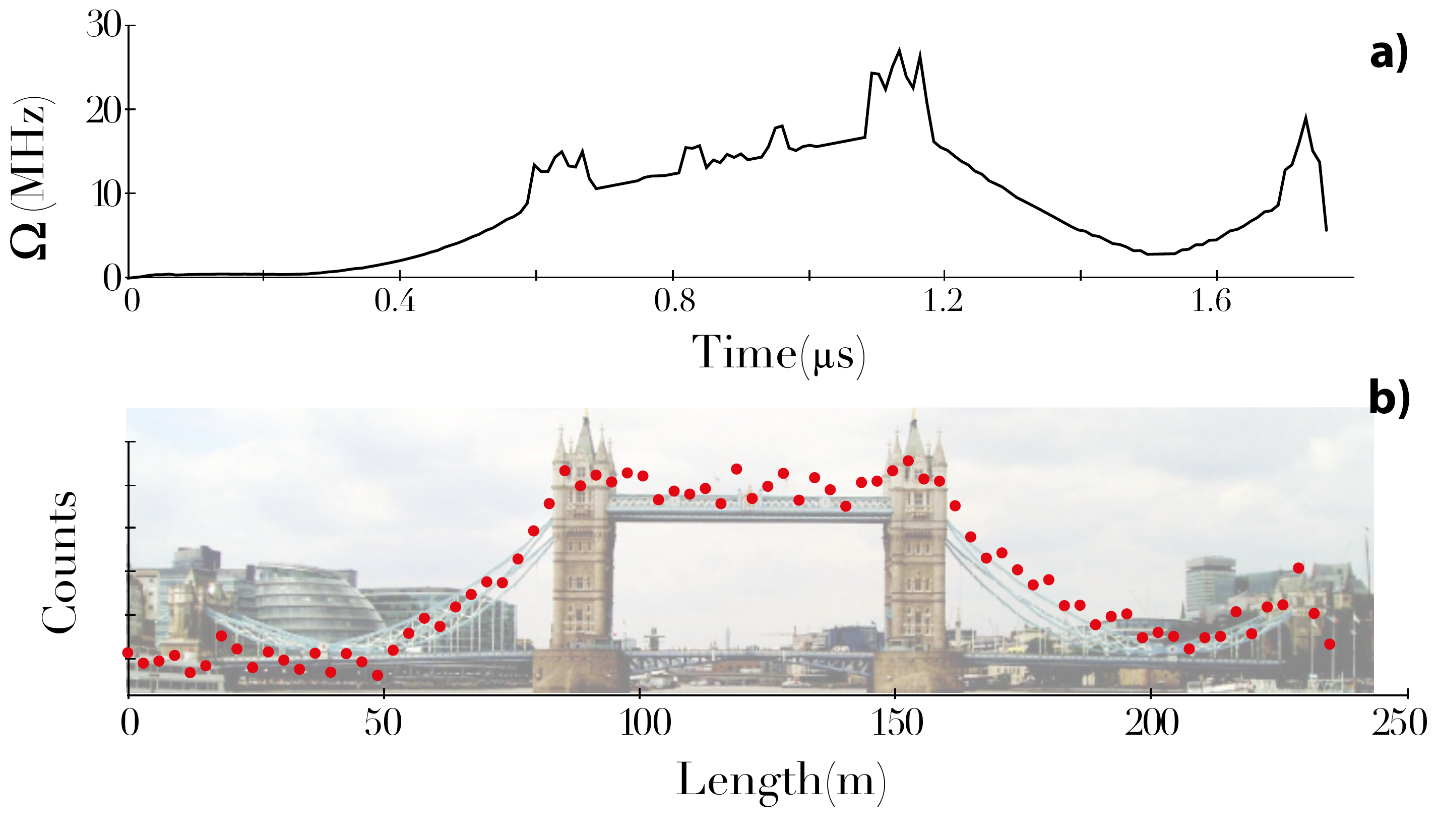}
\caption{(a) Driving pulse applied to an atom to obtain a photon with the shape shown below. The actual pulse was significantly smoothed due to the limited AOM bandwidth. (b) Spatial profile of the photon's probability density, reconstructed from the measured detection-time histogram. This has been done assuming that the light propagates in a fiber with a refractive index of $n=2$. Any resemblance to iconic landmarks is intentional. }
\label{fig:tb}
\end{figure}

It is also possible to engineer the photon's shape in more interesting ways.  By tailoring the Rabi frequency of the driving laser it is possible to force the photon amplitude to follow, for example, the shape of Tower Bridge in London, shown in  Fig.\ref{fig:tb}.   The shaping is fundamentally limited by the atom-cavity coupling $g_{0}$ and cavity decay $\kappa$, however in practice this is limited by the, AOM bandwidth ($5\,$MHz) used to modulate the driving pulse's amplitude.   

The indistinguishability of the emitted photons can be tested using a Hong-Ou-Mandel (HOM) interferometer \cite{Hong:1987zz}.  By introducing an optical delay, two photons from successive pulses are overlapped on a beam-splitter. In the case where they are identical bosons they will coalesce and leave the BS through the same output port, and for a perfect source, simultaneous detection at detectors 1 and 2 should not occur. As the temporal resolution of the detectors ($350$ps) is much shorter than the photon length, one can also observe the beat of the two photons\cite{Legero:2004p2114}. In a non-temporally resolved HOM interferometer, the temporal overlap of the photons on the beam-splitter is varied and the resultant correlations plotted. In this time-resolved case, the photons are always perfectly overlapped and the correlations instead vary with the time difference between clicks on the two detectors $\delta\tau$.  The theoretical background is set out in \cite{Legero:2003gn,Legero:2006gf}.

\begin{figure}
\centering
\includegraphics[width=8.5cm]{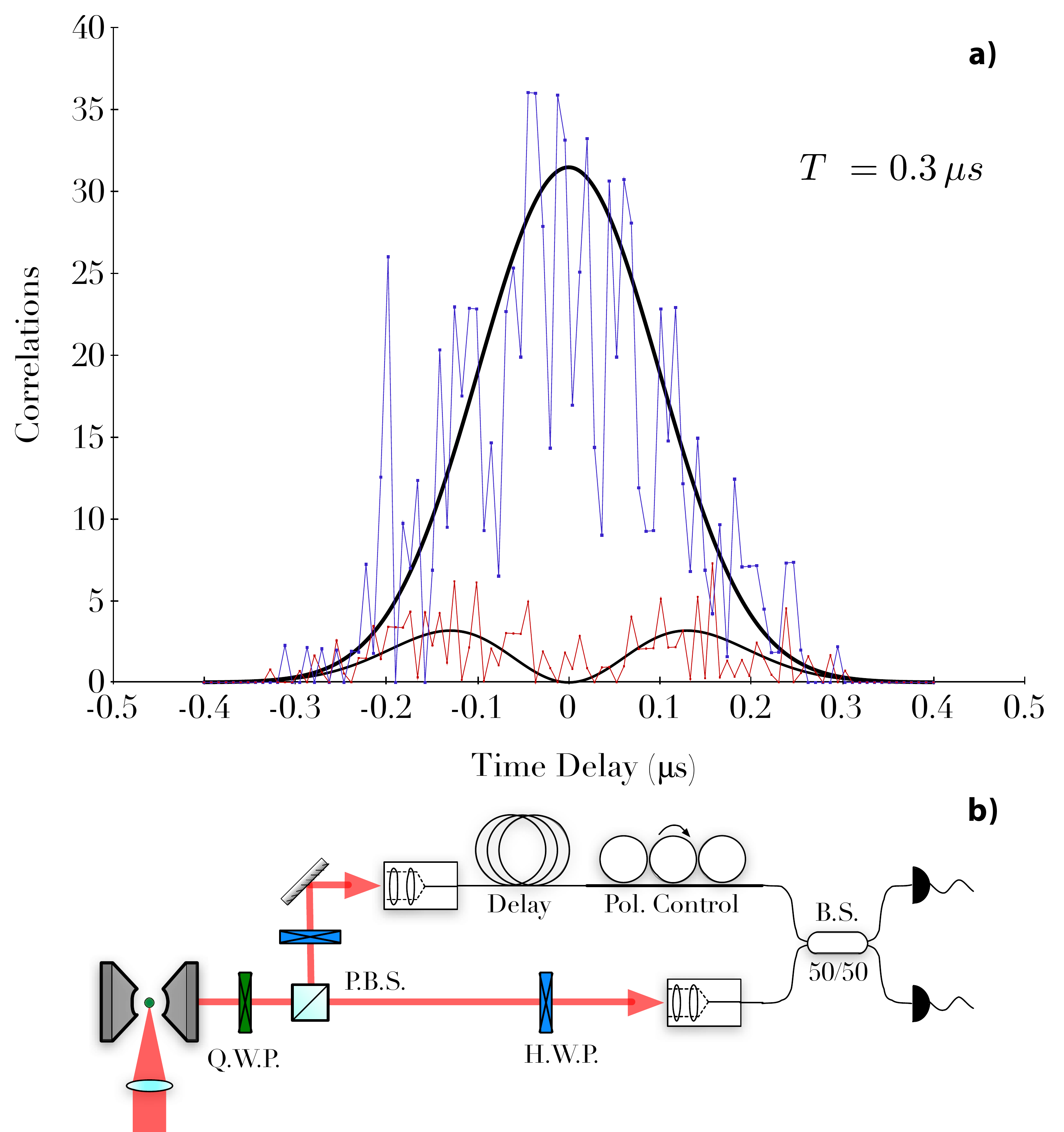}
\caption{(a) Cross correlation function of detectors 1 and 2, demonstrating 2-photon interference for photons overlapping spatially and temporally on a beamsplitter. Perpendicular polarized photons are shown in blue and parallel polarized photons in red. (b) The interferometer setup. A 200m long fiber is used to delay one photon such that it can be overlapped with the subsequently emitted one.  A fiber-based beamsplitter and polarisation optics are used to ensure near-perfect spatial and polarization overlap.}
\label{fig:HOM}
\end{figure}

 The interferometer is shown in Fig. \ref{fig:HOM} (a), and the cross correlation between the two detectors is shown in Fig. \ref{fig:HOM} (b) for parallel (red) and perpendicular (blue) polarizations. With parallel polarizations the two photons should interfere and correlations should not occur, whereas perpendicular polarizations do not interfere and a correlation function given by the convolution of the shapes of the two photons is observed.

As expected, the number of correlations for parallel polarizations is greatly reduced compared to the perpendicular case. In addition, a pronounced dip at $\delta\tau=0$ is visible, the width of which is governed by a characteristic `coherence time' $T=300\pm 40\,\mathrm{ns}$. This coherence time is engendered by a dephasing between the two photons, and is primarily limited by the stability of the pumping laser and stray magnetic fields.

We define the visibility $V_{2ph}$ as the reduction in the areas of the correlation curves between the completely distinguishable and completely indistinguishable case.
\begin{equation}
V_{2ph}=1-\frac{\int\!\Phi_{\parallel}(\tau)\,\mathrm{d}\tau}{\int\!\Phi_{\perp}(\tau)\,\mathrm{d}\tau}
\end{equation}
This is the time resolved equivalent of looking at the depth of the $\tau=0$ dip in non time-resolved HOM - the extra information is simply averaged away.   The overall extinction ratio for the photon is $V_{2ph}=0.87\pm0.05$.

In summary, we have demonstrated a cavity-based, deterministic single-photon source.  Emission rates of $1\/$MHz can be achieved, and the lack of disturbances caused by trapping fields means  the source exhibits efficiencies of over 60\%. Despite the lack of a trap, atoms remain inside the cavity mode for up to $100\,\mu$s giving enough time for QIP operations to be performed \cite{Himsworth:2011hn}.  The emitted photons show very strong anti-bunching and an indistinguishability leading to a HOM visibility of 87\%. We have also shown control over the shape of the photon's wavefunction, a requirement for an effective quantum memory \cite{Cirac:1997p1283,Specht:2011ir,Dilley:2011tf}.  The experimental arrangement is simple and thus more readily reproducible than similar sources and the source therefore shows great promise for testing the individual nodes from which a scalable quantum network will be composed.

\acknowledgements
This work was supported by the Engineering and Physical Sciences Research Council\linebreak (EP/E023568/1), the Deutsche Forschungsgemeinschaft (Research Unit 635),  and the EU through the RTN EMALI (MRTN-CT-2006-035369) and the QIP IRC.  We are grateful to D.\,Ljunggren and G.\,Langfahl-Klabes for their help in the early stages of this work. 

\bibliographystyle{apsrev4-1}
\bibliography{Library}

\end{document}